\documentclass[twocolumn,10pt]{IEEEtran}
\usepackage{epsfig, multirow, array, cite}
\usepackage{amssymb, amsmath, amsfonts, subfigure,color}
\newtheorem{lemma}{\textbf{Lemma}}
\newtheorem{corollary}{\textbf{Corollary}}
\newtheorem{theorem}{\textbf{Theorem}}

\allowdisplaybreaks

\begin{document}
\title{Caching Placement in Stochastic Wireless Caching Helper Networks:
Channel Selection Diversity via Caching}

\author{Seong Ho Chae,~\IEEEmembership{Member,~IEEE} and Wan Choi,~\IEEEmembership{Senior Member,~IEEE}
\thanks{A part of this paper will be presented at the IEEE Int. Workshop on Signal Processing Advanced in Wireless Commun. (SPAWC), Edinburgh, UK, Jul. 2016.}
\thanks{S. H. Chae and W.~Choi are with the School of Electrical Engineering,
Korea Advanced Institute of Science and Technology (KAIST), Daejeon, 34141, Korea (E-mail: shchae82@kaist.ac.kr, wchoi@kaist.edu).}
}

\maketitle

\begin{abstract}
Content delivery success in wireless caching helper networks depends mainly on cache-based channel selection diversity and network interference.
For given channel fading and network geometry,
both channel selection diversity and network interference dynamically vary
according to what and how the caching helpers cache at their finite storage space.
We study probabilistic content placement (or caching placement) to desirably control cache-based channel selection diversity and network interference in a stochastic wireless caching helper network, with sophisticated considerations of wireless fading channels, interactions among multiple users such as interference and loads at caching helpers, and arbitrary memory size.
Using stochastic geometry, we derive optimal caching probabilities in closed form to maximize the average success probability of content delivery and propose an efficient algorithm to find the solution in a noise-limited network. In an interference-limited network, based on a lower bound of the average success probability of content delivery, we find near-optimal caching probabilities in closed form to control the channel selection diversity and the network interference. We numerically verify that the proposed content placement is superior to other comparable content placement strategies.
\end{abstract}



\section{Introduction}
Recent evolution of mobile devices such as smart-phones and tablets has facilitated access to multi-media contents anytime and anywhere
but such devices result in an explosive data traffic increase.
The Cisco expects  by 2019 that these traffic demands will be grown up to 24.3 exabytes per month
and the mobile video streaming traffic will occupy almost 72\% of the entire
data traffic \cite{Cisco}.
Interestingly, numerous popular contents are asynchronously but repeatedly requested by many users
and thus substantial amounts of data traffic  have been redundantly generated over networks \cite{Caire}.
Motivated by this, caching or pre-fetching some popular video contents at the network edge
such as mobile hand-held devices or small cells (termed as \textit{local caching})
has been considered as a promising technique to alleviate the network traffic load.
As the cache-enabled edge node plays a similar role as a local proxy server with a small cache memory size,
the local wireless caching has the advantages of
i) reducing the burden of the backhaul by avoiding the repeated transmission of the same contents from the core network to end-users
and ii) reducing latency by shortening the communication distance.

In recent years, there have been growing interests in wireless local caching. The related research has focused mainly on i) femto-caching with cache-enabled small cells or access points (called as caching helpers) \cite{Caire2,Song15,SHChae15,Blaszczyszyn15,14Debbah,Hanzo15,Zhou15,ZChen15}, ii) device-to-device (D2D) caching with mobile terminals \cite{Kang14,Dimakis,Molisch,Andres,RuiZhang15,Afshang1,Afshang2},
and iii) heterogeneous cache-enabled networks \cite{Yang15,Rao15,Serbetci16}.
For these local caching networks,
varieties of content placements (or caching placements) were developed \cite{Caire2,Kang14,Song15,SHChae15,Blaszczyszyn15,Hanzo15,RuiZhang15,Rao15,Serbetci16}
and for given fixed content placement, the performance of cache-enabled systems
with different transmission or cache utilization techniques was investigated
\cite{14Debbah,Andres,Dimakis,Molisch,Yang15,Zhou15,Afshang1,Afshang2,ZChen15}.
Specifically,
content placement to minimize average downloading delay \cite{Caire2}
or average BER \cite{Song15} was proposed for fixed network topology.
In a stochastic geometric framework, various content placements were also proposed
either to minimize the average delay \cite{Hanzo15,RuiZhang15} and average caching failure probability \cite{Kang14}
or to maximize total hit probability \cite{Serbetci16}, offloading probability \cite{Rao15}.
However, these caching solutions were developed in limited environments; they discarded wireless fading channels and interactions among multiple users, such as interference and loads at caching helpers.

Recently, the content placement on stochastic geometry modeling of caching was studied in \cite{SHChae15,Blaszczyszyn15}.
A tradeoff between content diversity and cooperative gain according to content placement was  discovered well in \cite{SHChae15} but the caching probabilities were determined with numerical searches only. Moreover, in \cite{SHChae15}, cache memory size is restricted to a single content size and loads at caching helpers are not addressed.
The optimal geographical caching strategy to maximize the total hit probability was studied in cellular networks in \cite{Blaszczyszyn15}. However, only hit probability whether the requested content is available or not among the covering base stations was investigated. None of the previous works successfully addressed the channel selection diversity and interactions among multiple users
such as network interference and loads according to content placement.

Success of content delivery in wireless cache network depends mainly on two factors:
i) \textit{channel selection diversity gain} and ii) \textit{network interference}.
For given realization of nodes in a network, these two factors dynamically vary
according to what and how the nodes cache at their limited cache memory.
Specifically, if the more nodes store the same contents, they offer
the shorter geometric communication distance as well as the better small-scale fading channel for
the specific content request, which can be termed as channel selection diversity gain.
On the contrary, if the nodes cache all contents uniformly,
they can cope with all content requests but channel selection diversity gain cannot help being small. Moreover, according to content placement, the serving node for each content request dynamically changes, so the network interference from other nodes also dynamically varies. Thus, it might be required to properly control the channel selection diversity gain and network interference
for each content.

Recently, in \cite{Song15}, a tradeoff between content diversity and channel diversity was addressed in caching helper networks,
where each caching helper is capable of storing only \emph{one content}.
However, although pathloss and small-scale fading are inseparable in accurately modeling wireless channels,
the channel diversity was characterized with only small-scale fading and
the effects of pathloss and network interference depending on random network geometry were not well captured. In this context, we address the problem of content placement
with a more generalized model considering  pathloss, network interference according to random network topology based on stochastic geometry,
small-scale channel fading, and arbitrary cache memory size.
In this generalized framework, we develop an efficient content placement to desirably control cache-based channel selection diversity and network interference. The main contributions of this paper are summarized as follows.
\begin{itemize}
\item We model the stochastic wireless caching helper networks,
    where randomly located caching helpers store contents independently and probabilistically
    in their finite cache memory and each user receives the content of interest from the caching helper
    with the largest instantaneous channel power gain.
    Our framework generalizes the previous caching helper network models \cite{Song15,SHChae15}
    by  simultaneously considering small-scale channel fading, pathloss, network interference, and arbitrary cache memory size.

\item With stochastic geometry, we characterize the channel selection diversity gain
      according to content placement of caching helpers
      by deriving the cumulative distribution function of the smallest reciprocal of the channel power gain
      in a noise-limited network. We derive the optimal caching probabilities for each file in closed form to maximize the average content delivery success probability for given finite cache memory size, and propose an efficient algorithm to find the optimal solution.
\item In interference-limited networks, we derive a lower bound of the average content delivery success probability in closed form. Based on this lower bound with Rayleigh fading, we derive near-optimal caching probabilities for each content in closed form to appropriately control the channel selection diversity and the network interference depending on content placement.

\item Our numerical results demonstrate  that the proposed content placement is superior to other content placement strategies because the proposed method efficiently balances channel selection diversity and network interference reduction for given content popularity and cache memory size. We also numerically investigate the effects of the various system parameters, such as the density of caching helpers, Nakagami fading parameter, memory size, target bit rate, and user density, on the caching probability.
\end{itemize}

The rest of this paper is organized as follows.
In Section II, we describe the system model and performance metric considered in this paper.
We analyze the average content delivery success probability and desirable content placement of caching helpers in a noise- and interference-limited network in Sections III and IV, respectively.
Numerical examples to validate the analytical results and to investigate the effects of the system parameters
are provided in Section V. Finally, the conclusion of this paper is given in Section VI.

\section{System model and performance metric}
\begin{figure}[t!]
   \centerline{\psfig{figure=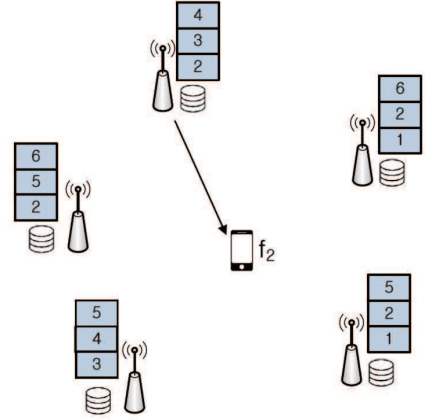,width=0.7\columnwidth} }
   \caption{System model}
   \label{Fig:system_model}
\end{figure}
We consider a downlink wireless video service network, where the caching helpers
are capable of caching some contents in their limited caching storage, as depicted in Fig. \ref{Fig:system_model}. We assume that all contents have the same size normalized to one for analytic simplicity\footnote{Unequal content sizes are not addressed in this paper but for unequal sized contents, each content can be
partitioned into small chunks of the same size. Each chunk can be treated as an individual content of the same size and then the analytic framework of this paper might be applicable.}.
The caching helpers are randomly located and modeled as
$2$-D homogeneous Poisson point process (PPP) with intensity $\lambda$. The caching helpers are equipped with a single antenna and
their cache memory size is $M$, so $M$ different contents can be cached at each helper since each content has unit size.
The total number of contents is $F(> M)$ and the set (library) of content indices is denoted as $\mathcal{F}=\{1,2,\cdots,F\}$.
The contents have own popularity\footnote{The contents have own popularity which is assumed to be perfectly known. However,  given time-varying content popularity in practical scenarios, incorporation of estimation errors of content popularity might be required and would be an interesting topic although this paper does not address it.}
and their popularity distributions are assumed to
follow the Zipf distribution as in literature \cite{Dimakis,SHChae15,Song15}:
\begin{align}
f_i=\frac{1/i^{\gamma}}{\sum_{j=1}^F 1/j^{\gamma}},~~\textrm{for}~~ i\in\mathcal{F},\label{Eqn:content_pop_dist}
\end{align}
where the parameter $\gamma(\geq 0)$ reflects the popularity distribution skewness.
For example, if $\gamma = 0$, the popularity of the contents is uniform.
The lower indexed content has higher popularity, i.e., $f_i\geq f_j$ if $i<j$.
Note that our content popularity profile is not necessarily confined to the Zipf distribution
but can accommodate any discrete content popularity distribution.
The users are also randomly located and modeled as $2$-D homogeneous Poisson point process (PPP) with intensity $\lambda_u$. Based on Slivnyak's theorem \cite{SGBook} that the statistics observed at a random point of a PPP $\Phi$
is the same as those observed at the origin in the process $\Phi\cup\{0\}$, we can focus on a reference user located at the origin, called a \emph{typical user}.

\begin{figure}[t!]
\centerline{\psfig{figure=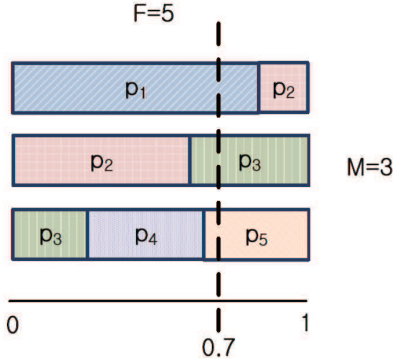,width=0.65\columnwidth} }
\caption{An example of the probabilistic caching method \cite{Blaszczyszyn15} when $M=3$ and $F=5$}
\label{Rev:Caching_explain}
\end{figure}

In this paper, we adopt \emph{random content placement} where
the caching helpers independently cache content $i$ with probability $p_i ~(0\leq p_i\leq 1)$ for all $i \in \mathcal{F}$.
According to the caching probabilities (or policies) $\{p_i\}$,
each caching helper randomly builds a list of up to $M$ contents to be cached by the probabilistic content caching method proposed in \cite{Blaszczyszyn15}.
Fig. 2 presents an example of the probabilistic caching method \cite{Blaszczyszyn15}
and illustrates how a caching helper randomly chooses $M$ contents to be cached among total $F$ contents
according to the caching probability $\{p_i\}$
when the cache memory size is $M=3$ and total number of contents is $F=5$.
In this scheme, the cache memory of size $M$ is equally divided into $L$ ($\leq M$) blocks of unit size.
Then, starting from content 1, each content sequentially fills the $M$ discontinuous memory blocks by the amount of $p_i$ from the first block. If a block is filled up in the filling process of content $j$, the remaining portion of content $j$ continuously fills the next block. Then, we select a random number within  $[0,1]$ and the contents at the position specified by the random number in each block are selected. Because one content is selected from each block by the selected random number, total $L$ ($\leq M$) contents can be selected in a probabilistic sense according to $\{p_i\}$.
In this way, in Fig. \ref{Rev:Caching_explain}, the contents $\{1,3,5\}$ are chosen to be cached.

The contents selected in a probabilistic sense at each helper are cached in advance by either its request or overhearing. The caching helpers storing content $i$ can be modeled as independent PPP with intensity
$\lambda_i(\triangleq p_i\lambda)$  and the locations of the caching helpers storing content $i$ can be represented by $\Phi_i=\{x_{i,k}\}$ where $k\in\mathbb{N}$.
The typical user requests one among $F$ contents according to the content popularity $f_i$;
the content with a higher popularity is requested with higher likelihood. If the typical user requests content $i$ and selects a serving helper to maximize the instantaneous channel power gain among the helpers storing content $i\in\mathcal{F}$, the received signal power becomes \begin{align}
\max_{x\in\Phi_i} P|h_x|^2|x|^{-\alpha},
\end{align}
where  $P$ is the transmit power of a caching helper, $h_x$ and $|x|$ denote the channel fading coefficient and the distance from the typical user to the caching helper located at $x$, respectively,
and $\alpha(>2)$ is the path loss exponent.

For each content $i$, we denote a set of the reciprocals of the channel power gains from $\Phi_i$ to the typical user in ascending order
as $\Xi_i = \left\{\xi_{i,k}=\frac{r_{i,k}^{\alpha}}{|h_{i,k}|^2}, k\in\mathbb{N}\right\}$, where
$\xi_{i,1}\leq \xi_{i,2}\leq \cdots$. The notation $r_{i,k}$ and $h_{i,k}$ represent the distance and the channel fading coefficient from the typical user to the caching helper with the $k$-th smallest reciprocal channel power gain
among the caching helpers storing content $i$, respectively.
Note that the caching helper with the largest instantaneous channel power gain
is equivalent to that with the smallest reciprocal of the channel power gain (i.e., $\xi_{i,1}$).
Assuming Gaussian signaling and time/frequency resource sharing among the users associated with the same caching helper, the mutual information
between the typical user requesting content $i$ and its serving caching helper is
\begin{align}
R_i=\frac{1}{N_i}\log_2\left(1+\frac{P}{\xi_{i,1}\left(\sigma^2+J_i(\xi_{i,1})\right)}\right),\label{R_i_with SINR}
\end{align}
where $N_i$ is the load of the serving caching helper,
$\sigma^2$ is the noise power variance, and $J_i(\xi_{i,1})$ is the  interference received at the typical user, given by
\begin{align}
J_i(\xi_{i,1})=\sum_{y\in\Phi_i^c} P|h_y|^2|y|^{-\alpha}
+\sum_{z\in\Xi_i\setminus\xi_{i,1}} Pz^{-1},
\end{align}
where
$\Phi_i^c(\triangleq\Phi\setminus \Phi_i)$ is the set of caching helpers which do not cache content $i$
in their cache memory.
The small-scale channel fading terms of the desired link and the interfering links follow the independent Nakagami-m distributions with parameters $m_D$ and $m_I$, respectively.

Similar to \cite{14Debbah,SHChae15},
we define the average content delivery success probability as a performance metric to properly account for the success events of content delivery
 as
\begin{align}
P_s = \sum_{i=1}^F f_i\cdot\mathbb{P}\left[R_i\geq \rho_i\right],\label{Def:FTSP}
\end{align}
where $f_i$ is the content requesting probability and $\rho_i$ is the target bit rate of content $i$ [bits/s/Hz]
to successfully support the real-time video streaming service of content $i$
without playback delay.

\section{Proposed Content Placement in Noise-limited Network}

In this section, in order to investigate how channel selection diversity affects the optimal caching solution, we first consider a noise-limited network; when the number of active users is much smaller than the number of caching helpers,
the impact of interference is negligible compared to the noise power
and the typical user can be served without resource sharing with other users.

In noise-limited networks, assuming Gaussian signaling, the mutual information between the typical user
requesting content $i$ and its serving helper is obtained as
\begin{align}
R_i=\log_2\left(1+\frac{1}{\xi_{i,1}}\frac{P}{\sigma^2}\right)=\log_2\left(1+\frac{\eta}{\xi_{i,1}}\right),
\end{align}
where $\eta=P/\sigma^2$ is the signal-to-noise ratio (SNR).

\subsection{Analysis of content delivery success probability}
The power gain distribution of a Nakagami-$m_s$ fading channel is given by
\begin{align}
f(x;m_s)&=\frac{m_s^{m_s}}{\Gamma(m_s)}x^{m_s-1}\exp\left(-m_s x\right),
\end{align}
where $\Gamma(t)=\int_0^{\infty}x^{t-1}e^{-x}dx$ is the gamma function,
$x\geq 0$, and $m_s \left( \geq \frac{1}{2} \right)$ is the fading parameter for link $s$
where $s\in \{D,I\}$ represents either the desired link ($D$) or the i.i.d. interfering links ($I$).
If $m_s=1$, the power gain distribution follows the exponential distribution corresponding to Rayleigh fading.
For $m_s\rightarrow \infty$, the channel is a deterministic channel.

When the typical user receives content $i$ from the caching helper with
the smallest reciprocal of the channel power gain (i.e., the largest channel power gain),
the cumulative distribution function (CDF) of the smallest reciprocal of the channel power gain (i.e., $\xi_{i,1}$) is derived in Lemma 1.
\begin{lemma}
The CDF of the smallest reciprocal of the channel power gain, $\xi_{i,1}$, in a Nakagami-$m_D$ fading channel
is given by
\begin{align}
F_{\xi_{i,1}}(\xi)=1-\exp\left(-\kappa p_i\xi^{\delta}\right),
\end{align}
where $\kappa=\pi\lambda\frac{\Gamma(\delta+m_D)}{m_D^{\delta}\Gamma(m_D)}$ and $\delta=\frac{2}{\alpha}$.
\end{lemma}
\vspace{0.05in}

\begin{IEEEproof}
For $i\in\mathcal{F}$, let $\Psi_i=\{r_{i,k}^{\alpha}(=|x_{i,k}|^{\alpha}),~k\in\mathbb{N}\}$ be the
path losses between the typical user and the caching helpers caching content $i$.
From the mapping theorem [Theorem 2.34, \cite{Book}],~
$\Psi_i$ is a non-homogeneous PPP and its intensity function is given by
\begin{align}
\lambda_{\Psi_i}(x)=p_i\lambda\pi\delta x^{\delta-1},~x\in\mathbb{R}^{+},
\end{align}
where $\delta = 2/\alpha$.
Note that $\{\Psi_i\}, \forall i\in\mathcal{F}$ are also mutually independent  due to independence among $\{\Phi_i\}$.
Using the displacement theorem [Theorem 2.33, \cite{Book}],
we can also derive the intensity function of $\Xi_i = \left\{\xi_{i,k}=\frac{r_{i,k}^{\alpha}}{|h_{i,k}|^2}\right\}$
for a general Nakagami-$m_D$ fading channel as
\begin{align}
\lambda_{\Xi_i}(y) = p_i\lambda\pi\delta y^{\delta-1}\frac{\Gamma(\delta+m_D)}{m_D^{\delta}\Gamma(m_D)},~y\geq 0.
\label{Eqn:Intensity_of_Xi_i}
\end{align}
Since the PPP of $\Phi_i$ is transformed by the displacement and mapping theorems,
$\Xi_i$ is also a PPP \cite{Book}.
Therefore, the CDF of $\xi_{i,1}$ is obtained as
\begin{align}
F_{\xi_{i,1}}(\xi) &= \mathbb{P}\left(\xi_{i,1}<\xi\right),\\
& = 1-\mathbb{P}(\Xi_i[0,\xi)=0)\\
&=1-\exp\left(-\int_0^{\xi} \lambda_{\Xi_i}(y)dy\right)\\
&=1-\exp(-\kappa p_i\xi^{\delta})
\end{align}
where $\Xi_i[0,\xi)$ denotes the number point of $\Xi_i$ in a circle with a radius $\xi$
and $\kappa=\pi\lambda\frac{\Gamma(\delta+m_D)}{m_D^{\delta}\Gamma(m_D)}$.
\end{IEEEproof}
\vspace{0.05in}

\emph{Remark:} As $p_i\lambda$ or $\mathbb{E}[|h_{i,k}|^{2\delta}]=\frac{\Gamma(\delta+m_D)}{m_D^{\delta}\Gamma(m_D)}$ increases, the CDF of $\xi_{i,1}$ grows faster to 1  because the intensity $\lambda_{\Xi_i}$ of PPP  $\Xi_i$  is proportional to them.
In other words, as the number of caching helpers that are storing the content of interest
and accessible by the typical user increases or the small-scale fading channel becomes more deterministic,
the intensity of PPP $\Xi_i$ representing the reciprocal channel power gains grows and thus the smallest reciprocal $\xi_{i,1}$ becomes smaller.
Especially, for given $\lambda$ and $m_D$, the largest channel power gain (i.e., $1/\xi_{i,1}$) grows as $p_i$ increases, which implies an increase of the \emph{channel selection diversity gain} according to the content placement.

\begin{figure}[!t]
   \centerline{\psfig{figure=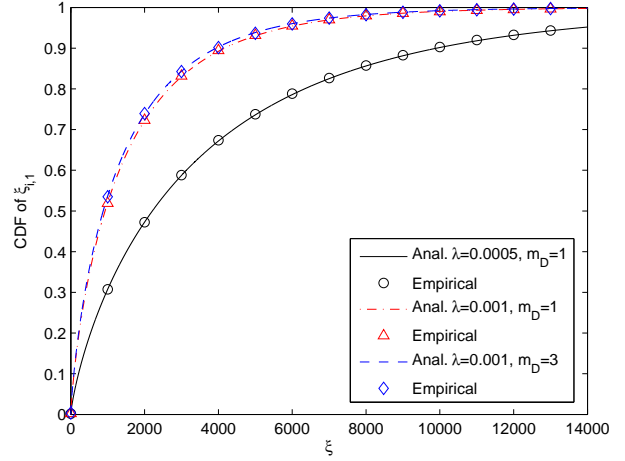,width=1.0\columnwidth} }
   \caption{The CDF of the smallest reciprocal of the channel power gain for content $i$ when $\alpha=2.5$ and $p_i=1$}
   \label{Fig:CDF}
\end{figure}

Fig. \ref{Fig:CDF} validates the accuracy of Lemma 1 for varying $\lambda$ and $m_D$.
The CDF of  $\xi_{i,1}$ increases faster to 1 as either $\lambda$ or $m_D$ increases. However, the CDF of  $\xi_{i,1}$ depends more on  $\lambda$ than on $m_D$, so  optimal caching probabilities are affected more by the density of caching helpers than channel fading.

From Lemma 1, the average success probability for content delivery is derived in the following theorem.
\begin{theorem}
When the typical user receives content $i$ from the caching helper with the largest instantaneous channel power gain,
the average success probability for content delivery $P_s$ in a Nakagami-$m_D$ fading channel  is obtained as
\begin{align}
P_s = 1-\sum_{i=1}^F f_i e^{-\kappa p_i\left(\frac{\eta}{2^{\rho_i}-1}\right)^{\delta}},\label{Eqn:FTSP}
\end{align}
where $\kappa=\pi\lambda\frac{\Gamma(\delta+m_D)}{m_D^{\delta}\Gamma(m_D)}$, $\delta=\frac{2}{\alpha}$,
$\eta=\frac{P}{\sigma^2}$, and $\rho_i$ is the target bit rate of content $i$.
\end{theorem}

\begin{IEEEproof}
\begin{align}
\mathbb{P}\left[R_i \geq\rho_i \right]
&=\mathbb{P}\left[\log_2\Big(1 + \frac{\eta}{\xi_{i,1}} \Big)\geq \rho_i\right]\\
&=\mathbb{P}\left[\xi_{i,1} \leq \frac{\eta}{2^{\rho_i} - 1}\right]\\
&=F_{\xi_{i,1}}\left(\frac{\eta}{2^{\rho_i} - 1}\right)\\
&=1-e^{-\kappa p_i\left(\frac{\eta}{2^{\rho_i}-1}\right)^{\delta}},\label{Eqn:Success_prob}
\end{align}
where \eqref{Eqn:Success_prob} is obtained from Lemma 1. Substituting \eqref{Eqn:Success_prob} into \eqref{Def:FTSP}, we obtain \eqref{Eqn:FTSP}.
\end{IEEEproof}

From Lemma 1, we know that the channel selection diversity gain for a specific content increases as the number of caching helpers storing the content increases, i.e., $p_i$ increases. However, due to limited memory space $M$, i.e., the constraint $\sum_{i=1}^F p_i \leq M$, storing the same content at more caching helpers ($p_i$ increases) loses the chance of storing the other contents and the corresponding channel diversity gains.

Therefore, for given finite memory size and content popularity,
the average success probability of content delivery can be maximized by
controlling the channel selection diversity gains for each content.
This can be achieved by optimally determining caching probabilities in random content placement. Consequently, the corresponding optimization problem can be formulated as
\begin{align}
\mathbf{P1:}~~\{p_i^{\star}\}
= \arg\max_{\{p_i\}} &\sum_{i=1}^F f_i\left[1- e^{-\kappa p_i\left(\frac{\eta}{2^{\rho_i}-1}\right)^{\delta}}\right],\\
=\arg\min_{\{p_i\}} &\sum_{i=1}^F f_ie^{-\kappa p_i\left(\frac{\eta}{2^{\rho_i}-1}\right)^{\delta}},\\
\textrm{subject to}~&\sum_{i=1}^F p_i \leq M,\label{Constraint1}\\
& p_i\leq 1~~~~\forall i \in \mathcal{F},\label{Constraint2}\\
& p_i\geq 0~~~~ \forall i \in \mathcal{F}.
\end{align}

\subsection{Optimal caching probabilities}\label{Section:opt}
In this subsection, we derive the optimal solution of Problem $\mathbf{P1}$, the optimal caching probabilities, in closed form. For each $i$, the function $g_i(p_i) \triangleq e^{-\kappa p_i\left(\frac{\eta}{2^{\rho_i}-1}\right)^{\delta}}$
is convex with respect to $p_i$ since $\frac{d^2}{d^2p_i}g_i(p_i)\geq 0$.
Since a weighted sum of convex functions is also convex function, Problem $\mathbf{P1}$ is a constrained convex optimization problem and thus a unique optimal solution exists. The Lagrangian function of problem $\mathbf{P1}$ is
\begin{align}
&\mathcal{L}(\{p_i\},\omega,\{\mu_i\})\nonumber\\
&=\sum_{i=1}^F f_ie^{-\kappa p_iT_i}\!+\omega\left(\sum_{i=1}^F p_i-M\!\right)\!+\!\sum_{i=1}^F\mu_i\left(p_i-1\right),
\end{align}
where $T_i\triangleq\left(\frac{\eta}{2^{\rho_i}-1}\right)^{\delta}$ is a constant, $\omega$ and $\mu_i$
are the nonnegative Lagrangian multipliers for constraints \eqref{Constraint1} and \eqref{Constraint2}.

After differentiating $\mathcal{L}(\{p_i\},\omega,\{\mu_i\})$ with respect to $p_i$, we can obtain
the necessary conditions for optimal caching probability, i.e., \textit{Karush-Kuhn-Tucker}(KKT) condition as follows:
\begin{align}
&\frac{\partial \mathcal{L}(\{p_i\},\omega,\{\mu_i\})}{\partial p_i}=-f_i\kappa T_ie^{-cp_iT_i}+\omega+\mu_i\geq 0,\label{Const1}\\
&\left\{-f_i\kappa T_ie^{-\kappa p_iT_i}+\omega+\mu_i\right\}p_i=0,\label{Const2}\\
&\omega\left(\sum_{i=1}^Fp_i-M\right)=0,\label{Const3}\\
&\mu_i\left(p_i-1\right)=0.\label{Const4}
\end{align}
From the constraint in \eqref{Const2},  the optimal caching probabilities are given by
\begin{align}
\!\!p_i(\omega,\mu_i)
&= \left[\frac{1}{\kappa T_i}\log\left(\frac{f_i\kappa T_i}{\omega+\mu_i}\right)\right]^{+}\\
&= \frac{1}{\kappa T_i}\left[\log\left(f_i\kappa T_i\right)-\log\left(\omega \!+\! \mu_i\right)\right]^{+},
~\forall i \!\in\! \mathcal{F},\label{Eqn:opt}
\end{align}
where $[z]^{+}=\max\{z,0\}$. The caching probability of content $i$ grows as the content popularity $f_i$ becomes large, but is regulated by the term of $\log (w +\mu_i)$.
For the constraint in \eqref{Const3}, $\omega$ is not necessarily zero because the optimal solution should always satisfy
$\sum_{i=1}^Fp_i=M$. Based on the KKT conditions in \eqref{Const1}-\eqref{Const4},
Lagrangian multipliers $\omega$ and $\mu_i$ range, according to $p_i$, as \eqref{Multiplier_range},
which is placed at the top of next page.

\begin{figure*}[!t]
\begin{align}
\left\{
\begin{array}{lll}
\omega\leq f_i\kappa T_ie^{-\kappa T_i},& \mu_i=\left[f_i\kappa T_ie^{-\kappa T_i}-\omega\right]^{+}  &~~\textrm{for}~~p_i=1,\\
f_i\kappa T_ie^{-\kappa T_i}<\omega<f_i\kappa T_i,&\mu_i=0~ &~~\textrm{for}~~0<p_i<1,\\
\omega \geq f_i\kappa T_i,&\mu_i=0 ~ &~~\textrm{for}~~p_i=0. \label{Multiplier_range}
\end{array}
\right.
\end{align}
\hrulefill
\end{figure*}


\eqref{Multiplier_range} reveals that the caching probability $p_i$ is determined
according to Lagrangian multiplier $\omega$ only since $\mu_i$ is a function of $\omega$; if $\omega\leq\min\{l_1,\cdots,l_F\}$ where $l_i=f_i\kappa T_ie^{-\kappa T_i}$, then $p_i=1 ~\forall i \in \mathcal{F}$ and thus $\sum_{i=1}^Fp_i(\omega,\mu_i)=F$. If $\omega\geq\max\{u_1,\cdots,u_F\}$ where $u_i=f_i\kappa T_i$, then $p_i=0 ~\forall i \in \mathcal{F}$ and thus  $\sum_{i=1}^Fp_i(\omega,\mu_i)=0$. When $\min\{l_1,\cdots,l_F\}\leq \omega \leq \max\{u_1,\cdots,u_F\}$,
$\sum_{i=1}^Fp_i(\omega,\mu_i)$ is bounded by $0\leq \sum_{i=1}^Fp_i(\omega,\mu_i) \leq F$ since $\sum_{i=1}^Fp_i(\omega,\mu_i)$ is decreasing with respect to $\omega$. Therefore, using the fact that
$\sum_{i=1}^Fp_i(\omega^{\star},\mu_i^{\star})=M$ for the optimal $\omega^{\star}$,  one-dimensional bisection search can find the optimal $\omega^{\star}$ and the corresponding $\{p_i(\omega^{\star},\mu_i^{\star})\}$ given by
\begin{align}
p_i^{\star} = \min\left([p_i(\omega^{\star},\mu_i^{\star})]^{+},1\right),~\forall i\in\mathcal{F}.\label{Opt_sol_noise}
\end{align}  The proposed algorithm to find the optimal caching probabilities $\{p_i(\omega^{\star},\mu_i^{\star})\}$ is presented in Algorithm 1.
Consequently, the content delivery success probability maximized with $\{p_i(\omega^{\star},\mu_i^{\star})\}$ becomes
\begin{align}
P_s^{\star} = \sum_{i=1}^F f_i \left[1- e^{-\kappa p_i^{\star}\left(\frac{\eta}{2^{\rho_i}-1}\right)^{\delta}}\right].
\end{align}
\vspace{0.1in}

\begin{figure}[!t]
\renewcommand{\arraystretch}{0.95}
\begin{tabular}{l}
\hline
\textbf{Algorithm 1.} A bisection method for finding $p_i^{\star}$\\
\hline
~1: $a \leftarrow \min\{l_1,\cdots,l_F\}$,\\
~~~~\hspace{0.01in}$b \leftarrow \max\{u_1,\cdots,u_F\}$~
$\triangleright~a,b$: two initial boundaries\\
~~~~$\omega \leftarrow \frac{a+b}{2}$ \hspace{0.65in}~~~~$\triangleright~\omega$: Initial guess of $\omega$\\
~2: \textbf{for} $i=1,\cdots,F$\\
~3: \hspace{0.2in} $\mu_i\leftarrow\left[l_i-\omega\right]^{+}$\\
~4: \hspace{0.2in} \textbf{Compute} \eqref{Eqn:opt}, i.e., $p_i(\omega,\mu_i)$\\
~5: \textbf{end}\\
~6: \textbf{while} $|\sum_{i=1}^{F} p_i(\omega,\mu_i)-M|\geq\epsilon$ \textbf{do}\\
    \hspace{0.2in} $\triangleright~\epsilon$: error tolerance level\\
~7: \hspace{0.2in} \textbf{if} $\sum_{i=1}^{F} p_i(\omega,\mu_i)>M$, \textbf{then} $a \leftarrow \omega$\\
~8: \hspace{0.2in} \textbf{else if} $\sum_{i=1}^{F} p_i(\omega,\mu_i)<M$, \textbf{then} $b \leftarrow \omega$\\
~9: \hspace{0.2in} \textbf{end if}\\
10: \hspace{0.2in} $\omega \leftarrow \frac{a+b}{2}$ \\
11: \hspace{0.2in} \textbf{Repeat:} Step 2 - 5\\
12: \textbf{end while}\\
13: $p_i^{\star}\leftarrow p_i(\omega,\mu_i)$~~\textbf{for}~$\forall i\in\mathcal{F}$\\
\hline
\end{tabular}
\end{figure}


\section{Proposed Content Placement in Interference-limited Network}

In the previous section, the cache-based channel selection diversity gain for each content has been highlighted
and the optimal caching probabilities to balance them were derived
without consideration of interference.
In this section, in the presence of network interference,
we derive near-optimal content placement and analyze the effects of network interference on the content placement.
We assume that the density of users is much higher than that of caching helpers, i.e., $\lambda_u\gg\lambda$, so the effect of noise is almost  negligible relative to interference.

\subsection{Analysis of content delivery success probability}
When the typical user receives content $i$ from the caching helper with the smallest reciprocal of the instantaneous channel power among the caching helpers storing content $i$,
the other caching helpers interfere with the typical user because they are assumed to serve other users.
Then, the received signal-to-interference ratio (SIR) at the typical user is represented as
\begin{align}
\gamma_i = \frac{P}{\xi_{i,1}\cdot J_i(\xi_{i,1})},\label{Eqn:SIR}
\end{align}
where $J_i(\xi_{i,1})$ is the interfering signal power and given by
\begin{align}
J_i(\xi_{i,1})=P\sum_{y\in\Phi_i^c} |h_y|^2|y|^{-\alpha}
+P\sum_{z\in\Xi_i\setminus\xi_{i,1}} z^{-1},
\end{align}
where
$\Phi_i^c(\triangleq\Phi\setminus \Phi_i)$ is a set of the caching helpers which do not cache content $i$
and $\Xi_i$ is a set of the reciprocals of the channel power gains from $\Phi_i$.
Note that the interfering signal power dynamically changes according to content placement of caching helpers since it is a function of $\xi_{i,1}$ and $\Phi_i^c$.
Therefore, optimal caching probabilities are expected to be obtained by
optimally controlling channel selection diversity and network interference for given content popularity and cache memory size.

\begin{figure}[t!]
\centerline{\psfig{figure=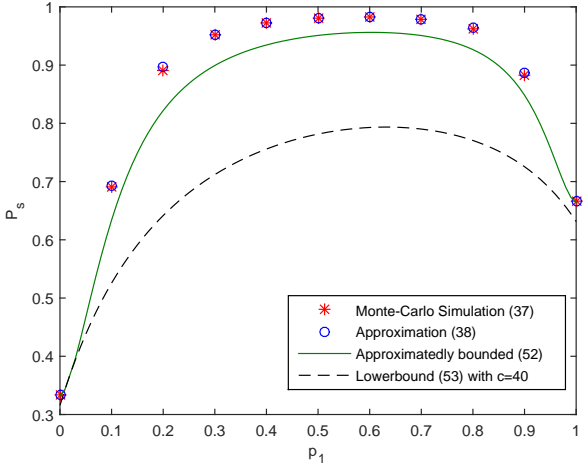,width=1.0\columnwidth} }
\caption{Average content delivery success probabilities of
Monte-Carlo simulation \eqref{Eqn:AVG_Ps_inter} and our approximations \eqref{Rev:5_1}, \eqref{Rev:5_2}, \eqref{Rev:5_3}
are compared versus $p_1=1-p_2$, when $\lambda = 10\mu$ (units/$m^2$), $\lambda_u = 20\mu$ (units/$m^2$), $\forall\rho_i=0.001$ (bits/s/Hz), $\gamma=1$, $m_D=m_I=1$, $M=1$, and $F=2$.}
\label{Fig:Approx_check}
\end{figure}

In interference-limited networks, the average success probability of content delivery in \eqref{Def:FTSP} is represented by
\begin{align}
P_s= \sum_{i=1}^F f_i \cdot
\mathbb{P}\left[\frac{1}{N_i^{\textrm{ins}}}\log_2\left(1+\frac{P}{\xi_{i,1}J_i(\xi_{i,1})}\right)\geq\rho_i\right],
\label{Eqn:AVG_Ps_inter}
\end{align}
where $N_i^{\textrm{ins}}$ is a random load of the tagged caching helper when an arbitrary user receives content $i$ from the caching helper with the largest instantaneous channel power gain.
To characterize \eqref{Eqn:AVG_Ps_inter}, we require both the probability mass function (PMF) of the load at the tagged caching helper and the SIR distribution when multiple contents are cached at each helper and the association is based on the instantaneous channel power gains.
However, unfortunately, the exact statistics of the required information are unavailable
because they are complicatedly determined by many interacting factors,
such as multiple cached contents, locations of caching helpers and users, content request of users,
instantaneous channel fading gains, etc.
Thus, the optimal caching probabilities to maximize \eqref{Eqn:AVG_Ps_inter} have to be found by
numerical searches of which complexity is prohibitively high for a huge number of contents. In this context, we propose near-optimal content placement to obtain some useful insights in interference-limited scenarios. To this end, we first approximate \eqref{Eqn:AVG_Ps_inter} with the average load of the tagged caching helper \cite{Singh1,Singh2,SHChae16} as
\begin{align}
P_s\approx\sum_{i=1}^F f_i \cdot
\mathbb{P}\left[\frac{1}{\bar{N}_i^{\textrm{ins}}}\log_2\left(1+\frac{P}{\xi_{i,1}J_i(\xi_{i,1})}\right)\geq\rho_i\right],\label{Rev:5_1}
\end{align}
where $\bar{N}_i^{\textrm{ins}}$ is the average load of the tagged caching helper when the user requests content $i$ to the caching helper with the largest instantaneous channel power gain.
The validity of approximation \eqref{Rev:5_1} is demonstrated in Fig. \ref{Fig:Approx_check}, where
red star and blue circle represent the Monte-Carlo simulation \eqref{Eqn:AVG_Ps_inter} and its approximation \eqref{Rev:5_1}, respectively.
This figure verifies that the approximation \eqref{Rev:5_1} is quite tight to \eqref{Eqn:AVG_Ps_inter} for arbitrary $p_1$.
Moreover, a lowerbound of \eqref{Rev:5_1} is obtained in the following theorem.

\begin{theorem}
When the typical user receives the requesting content from the caching helper with the smallest reciprocal of instantaneous channel power gain, the average success probability of content delivery is bounded below by
\begin{align}
P_s \geq \sum_{i=1}^F f_i \int_0^{\infty}
&\sum_{k=0}^{m_D-1}\frac{1}{k!}\left(-m_D P^{-1}\tau_i r^{\alpha}\right)^k\times\nonumber\\
&\frac{d^k}{ds^k}\mathcal{L}_{I_i}(s)|_{s=m_DP^{-1}\tau_i r^{\alpha}}f_{|x_i|}(r)dr,\label{Eqn:AVG_Ps_inter_lower}
\end{align}
where $\tau_i=2^{c\rho_i}-1$, $c ~(\geq 1)$ is a constant independent of $i$ and
makes the inequality hold for all ranges of $\{p_i\}$,
$m_D$ and $m_I$ are the Nakagami fading parameters of the desired and interfering links, respectively, and
\begin{align}
\mathcal{L}_{I_i}(s)
&=\exp\left(-2\pi\lambda\!\int_0^{\infty}\left[1-\frac{m_I}{(sPv^{-\alpha}+ m_I)^{m_I}}\right]vdv\right.\nonumber\\
&\left.+~2\pi p_i\lambda\int_0^r \!\left[1 -\frac{m_I}{(sPv^{-\alpha}+m_I)^{m_I}}\right]vdv\right),\\
f_{|x_i|}(r) &= 2\pi p_i\lambda r\exp\left(-\pi p_i\lambda r^2\right).
\end{align}
\end{theorem}
\begin{IEEEproof}
See Appendix \ref{appendixA}.
\end{IEEEproof}

Based on the lower-bounded average success probability of content delivery, we formulate an alternative optimization problem as
\begin{align}
\mathbf{P2:}~~\{\hat{p}_i^{\star}\}= &\arg\max_{\{p_i\}}~\eqref{Eqn:AVG_Ps_inter_lower},\nonumber\\
\textrm{subject to}~&\sum_{i=1}^F p_i \leq M,\nonumber\\
&p_i\leq 1~~~~\forall i \in \mathcal{F},\nonumber\\
&p_i\geq 0~~~~ \forall i \in \mathcal{F}. \label{Alternative_opt_prob}
\end{align}
Although it is still non-trivial to obtain the solution of this alternative optimization problem, fortunately, when $m_D=m_I=1$, i.e., a Rayleigh fading channel, the objective function (i.e., the lower bound of delivery success probability) becomes more tractable and sheds light on intuitively understanding the impacts of network interference on content placement. Therefore, in the following subsection, we focus on the case of $m_D=m_I=1$ (i.e., Rayleigh fading).

\subsection{Caching probabilities in Rayleigh fading channels}

\begin{corollary}
For Rayleigh fading channels (i.e., $m_D=m_I=1$), the lower-bound of delivery success probability
in \eqref{Eqn:AVG_Ps_inter_lower} is simplified as
\begin{align}
P_s \geq\sum_{i=1}^F f_i \cdot \frac{p_i}{\left(1-\tau_i^{2/\alpha}C_{\tau_i,\alpha}\right)p_i+\tau_i^{2/\alpha}C_{\alpha}},
\label{Eqn:AVG_FTSP_inter_lower2}
\end{align}
where $\tau_i=2^{c\rho_i}-1$, $C_{\alpha}=\frac{2\pi}{\alpha}\csc\left(\frac{2\pi}{\alpha}\right)$, $C_{\tau_i,\alpha} = \tau_i^{-2/\alpha}~_2F_1\left(1,\frac{2}{\alpha};1+\frac{2}{\alpha};-\frac{1}{\tau_i}\right)$
and $_2F_1(\cdot)$ is the Gauss hypergeometric function.
\end{corollary}

\begin{IEEEproof}
We omit the proof since it can be readily obtained by substituting $m_D=m_I=1$ in Theorem 2.
\end{IEEEproof}

With arbitrary cache memory size of $M$ at each helper, the alternative optimization problem
\textbf{P2} is rewritten as
\begin{align}
\mathbf{P3:}~\{\hat{p}_i^{\star}\}
\!=\arg\max_{\{p_i\}}&\sum_{i=1}^F f_i\frac{p_i}{\left(1 \!-\!\tau_i^{2/\alpha}C_{\tau_i,\alpha}\right)p_i \!+\tau_i^{2/\alpha}C_{\alpha}},\nonumber\\
\textrm{subject to}~&\sum_{i=1}^F p_i \leq M,\nonumber\\
& p_i\leq 1~~~~\forall i \in \mathcal{F},\nonumber\\
& p_i\geq 0~~~~ \forall i \in \mathcal{F}.\label{Prob:Interference}
\end{align}
Now we show that the objective function in \textbf{P3} is concave and optimization problem \textbf{P3} is also the constrained convex optimization problem.

If we define $g_i(p_i)$ as
\begin{align}
g_i(p_i) \triangleq \frac{p_i}{(1-A_i)p_i+B_i},
\end{align}
where $A_i=\tau_i^{2/\alpha}C_{\tau_i,\alpha}(>0)$ and $B_i=\tau_i^{2/\alpha}C_{\alpha}(> 0)$, its first derivative is $g_i'(p_i)=\frac{B_i}{\left[(1-A_i)p_i+B_i\right]^2}> 0$
because $B_i> A_i$ always holds and $(1-A_i)p_i+B_i> 0$ for $0\leq p_i\leq 1$.
Note that $0< A_i\leq 1$ for all $i$ because
\begin{align}
A_i& =\tau_i^{2/\alpha}C_{\tau_i,\alpha}=\tau_i^{2/\alpha}\int_0^{\tau_i^{-2/\alpha}}\frac{1}{1+u^{\alpha/2}}du \nonumber\\
&\leq\tau_i^{2/\alpha}\int_0^{\tau_i^{-2/\alpha}}1du=1,\\
A_i&=\tau_i^{2/\alpha}C_{\tau_i,\alpha}=\tau_i^{2/\alpha}\int_0^{\tau_i^{-2/\alpha}}\frac{1}{1+u^{\alpha/2}}du \nonumber\\
&\geq\tau_i^{2/\alpha}\int_0^{\tau_i^{-2/\alpha}}0du=0.
\end{align}
The second derivative of $g_i(p_i)$ is
$g_i''(p_i) = \frac{2B_i(A_i-1)}{\left[(1-A_i)p_i+B_i\right]^3}\leq 0$ and thus
$g_i(p_i)$ is a strictly increasing concave function.
Since a weighted sum of concave functions still satisfies
concavity, optimization problem  \textbf{P3} is a constrained convex optimization problem. Applying the same approach in Section \ref{Section:opt},
we obtain the optimal caching probability of problem \textbf{P3} as
\begin{align}
\!\!\! p_i(\omega,\mu_i) &\!=\!
\left[-\frac{B_i}{1 \!-\! A_i}\!+\!\sqrt{\!\frac{f_iB_i}{(1 \!-\! A_i)^2(\omega^{\star} \!+\! \mu_i^{\star})}}~\right]^{+}\!\!\!\!,~\forall i\!\in\!\mathcal{F},\\
&=\frac{1}{1 \!-\! A_i}\left[-B_i+\sqrt{\frac{f_iB_i}{\omega^{\star}\!+\!\mu_i^{\star}}}~\right]^{+}\!\!\!,~\forall i\!\in\!\mathcal{F},
\label{Eqn:opt2}
\end{align}
where Lagrangian multipliers $\omega$ and $\mu_i$ range, according to $p_i$, as \eqref{Multiplier_range_Inter},
which is placed at the top of next page.

\begin{figure*}[t!]
\begin{align}
\left\{
\begin{array}{lll}
\omega\leq \frac{f_iB_i}{\left(1-A_i+B_i\right)^2},~~~~&\mu_i=\left[\frac{f_iB_i}{\left(1-A_i+B_i\right)^2}-\omega\right]^{+}  & \textrm{for}~~p_i=1,\\
\frac{f_iB_i}{\left(1-A_i+B_i\right)^2}<\omega<\frac{f_i}{B_i},~~~&\mu_i=0~ &\textrm{for}~~0<p_i<1,\\
\omega \geq \frac{f_i}{B_i},~~~&\mu_i=0 ~ &\textrm{for}~~p_i=0. \label{Multiplier_range_Inter}
\end{array}
\right.
\end{align}
\hrulefill
\end{figure*}

Replacing \eqref{Eqn:opt} with \eqref{Eqn:opt2} and letting $l_i=\frac{f_iB_i}{\left(1-A_i+B_i\right)^2}$ and $u_i=\frac{f_i}{B_i}$ in Algorithm 1, we can find
the optimal $\omega^{\star}$ and $\mu_i^{\star}$ with one-dimensional bisection search and the corresponding near-optimal
caching probabilities $\{p_i (\omega^{\star}, \mu_i^{\star})\}$ given by
\begin{align}
\hat{p}_i^{\star}\!=\!
\min\!\left(\!\frac{1}{1\!-\!A_i}\left[-B_i \!+\!\sqrt{\frac{f_iB_i}{\omega^{\star}\!+\!\mu_i^{\star}}}~\right]^{+}\!\!,~1\right),~\forall i\!\in\!\mathcal{F}.
\label{Eqn:opt_sol_interlimited}
\end{align}

\emph{Remark:} Unlike noise-limited networks, the solution of content placement obtained in \eqref{Eqn:opt_sol_interlimited} is independent of the transmit power of caching helpers.
The caching probability is a function of $\alpha$, $f_i$ and $\tau_i=2^{c\rho_i}-1$.
In other words, the content placement is determined by the pathloss exponent, content popularity, and target bit rate.

\section{Numerical Results}\label{Section_Numerical}
In this section, we evaluate the average success probability of content delivery
to validate our analytical results in the previous sections.
We also examine how various system parameters,
such as $\mathrm{SNR}$, content popularity exponent ($\gamma$), Nakagami fading parameter ($m_D$ and $m_I$),
pathloss exponent ($\alpha$), density of caching helpers ($\lambda$), user density ($\lambda_u$),
maximum target content bit rate ($\rho_{\max}$), and cache memory size ($M$)
affect on caching probabilities. Unless otherwise stated,
the baseline setting of simulation environments is as follows:
$\gamma=1$, $F=10$, $M=3$, $m_D=m_I=1$, $\mathsf{SNR}$ = 20~(dB), $\alpha=3$,
$\lambda=0.05$ (units/$m^2$), $\lambda_u=0.002$ (units/$m^2$) and $\rho_{\max}=1$ (bits/s/Hz).
The target bit rate for each content is uniformly generated as $\rho_i\in\left(0,\rho_{\max}\right]$
and all simulation results are averaged over 10,000 realizations.

\begin{figure}[t!]
   \centerline{\psfig{figure=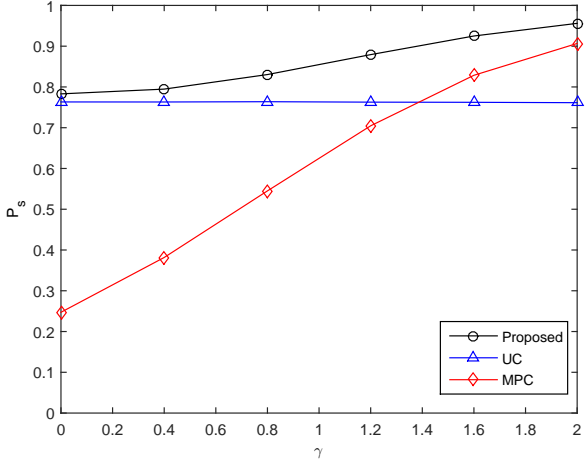,width=1.0\columnwidth} }
   \caption{Comparison of average content delivery success probabilities among the proposed content placement, UC, and MPC versus the content popularity exponent $\gamma$ where $F=20$ and $M=5$.}
   \label{Fig:caching_comparison}
\end{figure}

\subsection{Comparison among three different caching strategies}
Fig. \ref{Fig:caching_comparison} compares the average success probabilities of content delivery in a noise-limited network for three different content placement strategies; i) caching the $M$ most popular contents (MPC), ii) caching the contents uniformly (UC), and iii) proposed content placement found by Algorithm 1 (Proposed).
This figure demonstrates that the proposed content placement in \eqref{Opt_sol_noise} is superior to both
UC and MPC in terms of average success probability of content
delivery. MPC is closer to the proposed content placement than
UC for high $\gamma$, and vice versa for low $\gamma$.

\begin{figure}[t!]
   \centerline{\psfig{figure=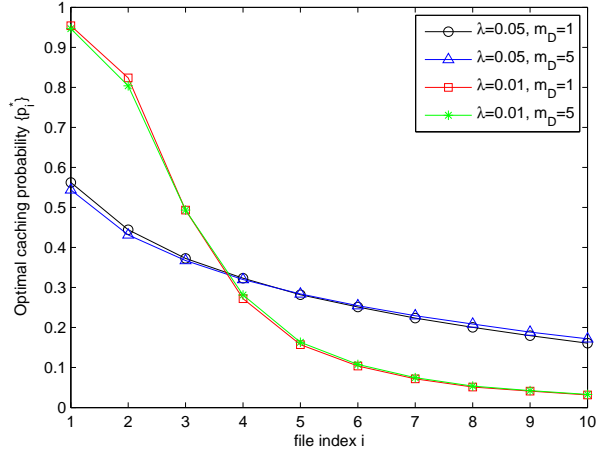,width=1.0\columnwidth} }
   \caption{Optimal caching probability of each content for varying $\lambda$ and $m_D$}
   \label{Fig:Opt_sol_lambda_m}
\end{figure}

\subsection{Effects of channel power gains}
For varying $\lambda$ and $m_D$, the optimal caching probability of each content $i$  in a noise-limited network
is plotted in Fig. \ref{Fig:Opt_sol_lambda_m}, where the lower index indicates the higher popularity, i.e., $f_i\geq f_j$ if $i< j$.
As $\lambda$ or $m_D$ increases, the optimal caching probability becomes more uniform. It implies that it would be beneficial to increase hitting probability for all contents instead of focusing on channel selection diversity for a few specific contents. This is because
channel power gains become higher as either the number of caching helpers increases or channels become more deterministic although channel selection diversity can be limited. This figure also exhibits that the optimal caching probability depends more on the geometric path loss than on small-scale fading, which matches the implication of Fig. \ref{Fig:CDF}.

\begin{figure}[t!]
   \centerline{\psfig{figure=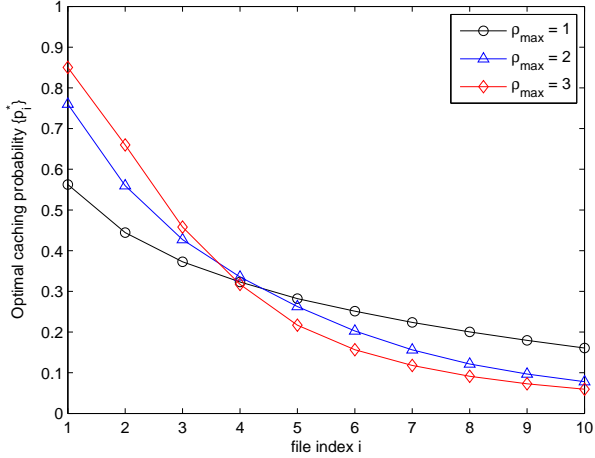,width=1.0\columnwidth} }
   \caption{Optimal caching probability of each content for varying maximum target bit rate $\rho_{\max}$}
   \label{Fig:Opt_sol_target_bit_rate}
\end{figure}

\subsection{Effects of target bit rate}
Fig. \ref{Fig:Opt_sol_target_bit_rate} shows the optimal caching probability of each content $i$
in a noise-limited network for varying maximum target bit rate $\rho_{\max}$. As $\rho_{\max}$ grows, the optimal caching probability becomes
biased toward caching the most popular contents. If
$\rho_{\max}$ is large, increasing channel selection diversity gains of the most popular contents is more beneficial to improve success probability of content delivery.

\subsection{Effects of cache memory size}
\begin{figure}[t!]
   \centerline{\psfig{figure=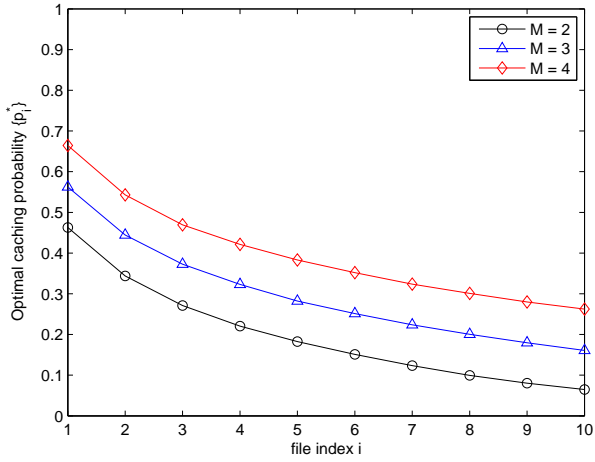,width=1.0\columnwidth} }
   \caption{Optimal caching probability of each content for varying $M$}
   \label{Fig:Opt_sol_M}
\end{figure}

In Fig. \ref{Fig:Opt_sol_M}, the optimal caching probability of each content $i$ in a noise-limited network is plotted for varying cache memory size $M$.
The optimal caching probabilities scale with the cache memory size $M$, but
they become more uniform as $M$ increases.
This is because less popular contents are accommodated in memory of larger size.

\subsection{Validation of the proposed near-optimal content placement}
\begin{figure}[t!]
\centerline{\psfig{figure=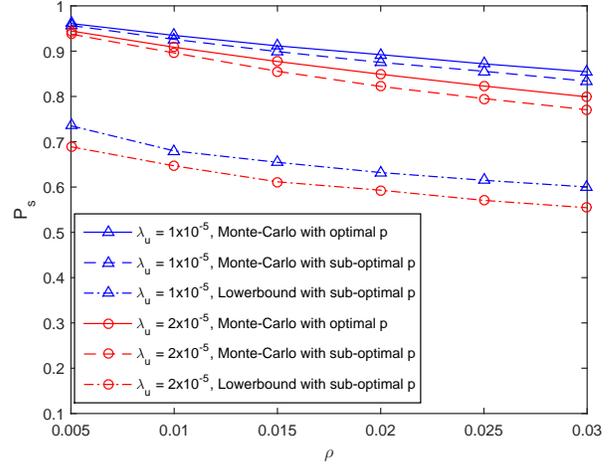,width=1.0\columnwidth} }
\caption{Comparison of average success probabilities of content delivery with
Monte-Carlo simulation with optimal $p^{\star}$, Monte-Carlo simulation with sub-optimal $\hat{p}^{\star}$,
and lowerbound with sub-optimal $\hat{p}^{\star}$
versus $\forall \rho_i=\rho$,  when $\lambda = 1\times10^{-5}$ (units/$m^2$), $M=1$, and $F=2$.}
\label{Fig:Ps_with_opt_and_subopt}
\end{figure}

Fig. \ref{Fig:Ps_with_opt_and_subopt} compares the average success probabilities of content delivery
with optimal $p^{\star}$ obtained from \eqref{Eqn:AVG_Ps_inter} by brute-force searches, with the proposed sub-optimal $\hat{p}^{\star}$ obtained from \textbf{P3}, and the lower bound \eqref{Eqn:AVG_Ps_inter_lower} with the sub-optimal $\hat{p}^{\star}$
versus $\forall \rho_i=\rho$, when $\lambda = 1\times 10^{-5}\mu$ (units/$m^2$), $\gamma=1$, $M=1$, and $F=2$.
For each $\rho$ and $\lambda_u$, the value of $c$ for a tighter lower bound is numerically found.
Since the content placement obtained from the lower bound is sub-optimal,
the average content delivery success probability with the sub-optimal $\hat{p}^{\star}$ is bounded below that with
optimal $p^{\star}$. Although there is a large gap between the lower bound in \eqref{Eqn:AVG_Ps_inter_lower} and $P_s$,
the gap between the average content delivery success probabilities with the optimal $p^{\star}$ and the proposed $\hat{p}^{\star}$
is small for an arbitrary target bit rate because  \eqref{Eqn:AVG_Ps_inter} and  \eqref{Eqn:AVG_Ps_inter_lower} have quite similar shapes. Consequently,
the proposed sub-optimal caching probability is close to optimal caching probability although the sub-optimal caching probability is found from the lower bound in \eqref{Eqn:AVG_Ps_inter_lower}.

\subsection{Comparison among three different caching strategies in interference-limited network}
\begin{figure}[t!]
\centerline{\psfig{figure=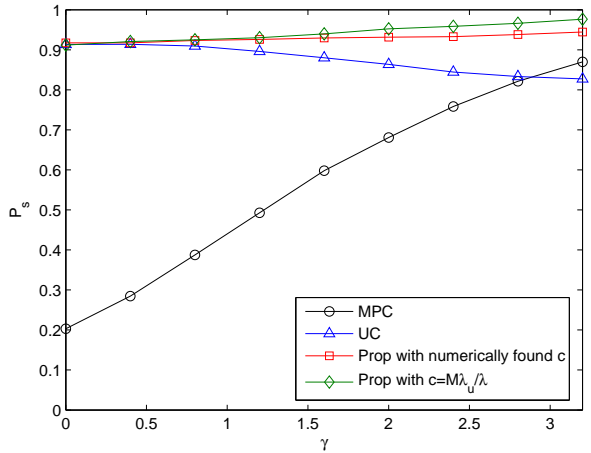,width=1.0\columnwidth} }
\caption{
Comparison of average content delivery success probabilities among the proposed content placement schemes with numerically found $c$ yielding a tight lower bound and with $c=\frac{M\lambda_u}{\lambda}$, UC, and MPC versus the content popularity exponent $\gamma$,
where $\lambda = 1\times 10^{-5}$ (units/$m^2$), $P$ = 20(dB), $F=5$, $M=1$, and $\forall \rho_i = 0.001$ (bits/s/Hz)}
\label{Fig:Inter_comparision}
\end{figure}

Fig. \ref{Fig:Inter_comparision} compares the average content delivery success probabilities
among the proposed content placement schemes with numerically found $c$ yielding a tight lower bound
and with $c=\frac{M\lambda_u}{\lambda}$, UC, and MPC versus the content popularity exponent $\gamma$.
Although the value of $c$ needs to be numerically found, any suboptimal solution even with the value $c$ which does not always satisfy the inequality in \eqref{Eqn:AVG_Ps_inter_lower} yields a lower average success probability of content delivery because of its suboptimality. From this fact, a suboptimal solution can be found by setting the value of $c$ to be the average load of a typical caching helper as $c= \frac{M\lambda_u}{\lambda}$ for simplicity.
Fig. \ref{Fig:Inter_comparision} demonstrates that
that both the proposed content placement schemes
with numerically found $c$ and $c=\frac{M\lambda_u}{\lambda}$
are superior to both UC and MPC in terms of average content delivery success probability for general $\gamma$.
The average content delivery success probability with $c= \frac{M\lambda_u}{\lambda}$ is quite similar to that with numerically found $c$ and outperforms UC and MPC.

\subsection{Effects of user density}
\begin{figure}[t!]
   \centerline{\psfig{figure=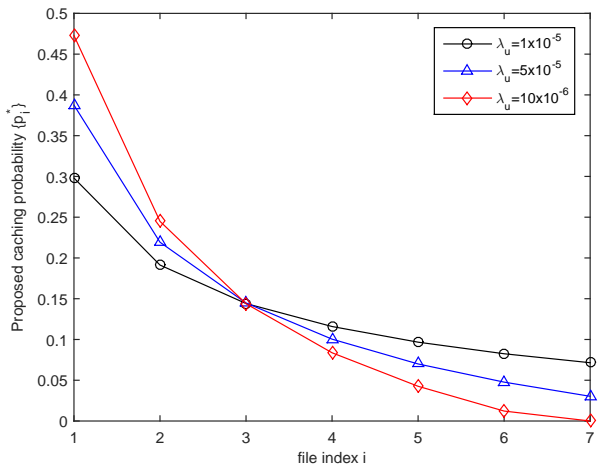,width=1.0\columnwidth} }
   \caption{The proposed sub-optimal caching probability of each content $i$
obtained by solving the convex optimization problem \eqref{Prob:Interference} for varying user density $\lambda_u$,
when $\lambda = 1\times 10^{-5}$ (units/$m^2$), $P$ = 20(dB), $M=1$, $F=7$, and $\forall \rho_i = 0.001$ (bits/s/Hz)}
   \label{Fig:Inter_opt_sol_user}
\end{figure}

In an interference-limited network, for varying user density $\lambda_u$,
the proposed caching probability of each content $i$
obtained by solving the convex optimization problem in \textbf{P3} is plotted in Fig. \ref{Fig:Inter_opt_sol_user},
where the value of $c$ yielding a tight lower bound is numerically found. As the user density $\lambda_u$
decreases, the optimal content placement tends to cache all contents with more uniform probabilities.

\section{Conclusions}
We studied probabilistic content placement to desirably control cache-based channel selection diversity and network interference in a wireless caching helper network, with specific considerations of path loss, small-scale channel fading, network interference according to random network topology based on stochastic geometry, and arbitrary cache memory size.
In a noise-limited case, we derived the optimal caching probabilities for each content in closed form in terms of the average success probability of content delivery and proposed a bisection based search algorithm to efficiently reach the optimal solution. In an interference-limited case, we derived a lower bound on the average success probability of content delivery. Then, we found the near-optimal caching probabilities in closed form in Rayleigh fading channels,  which maximize the lower bound. Our numerical results verified that the proposed content placement is superior to the conventional caching strategies because  the proposed scheme efficiently controls the channel selection diversity gain and the interference reduction. We also numerically analyzed the effects of various system parameters, such as caching helper density, user density, Nakagami $m$ fading parameter, memory size, target bit rate, and user density, on
the content placement.


\appendices

\section{Proof of Theorem 2}\label{appendixA}
Since the pathloss dominates the small-scale fading effects according to Lemma 1,  $\bar{N}_i^{\textrm{ins}}$ is approximated as the load of the tagged caching helper with the largest channel power gain averaged over fading (i.e., the load based on the association with long-term channel power gains), $\bar{N}_i^{\textrm{ins}}\approx \bar{N}_i$. Moreover, the received SIR with the association based on instantaneous channel power gains is larger than that with the association based on long-term channel power gains.
Accordingly, \eqref{Rev:5_1} can be further bounded below as
\begin{align}
\eqref{Rev:5_1}\geq \sum_{i=1}^F f_i\cdot
\mathbb{P}\left[\frac{1}{\bar{N}_i}\log_2\!\left(1\!+\!\frac{P|h_{x_i}|^2|x_i|^{-\alpha}}{I_i}\right)\!\geq\!\rho_i\right],\label{Rev:5_2}
\end{align}
where
\begin{align}
I_i = \sum_{y\in\Phi\setminus x_i} P|h_y|^2|y|^{-\alpha},\nonumber
\end{align}
which is also validated in Fig. \ref{Fig:Approx_check}, where
blue circle and green solid line represent \eqref{Rev:5_1} and \eqref{Rev:5_2}, respectively.

In case of $M=1$, a closed form expression of $\bar{N}_i$ is available as $\bar{N}_i=1+1.28\frac{f_i\lambda_u}{p_i\lambda}$ \cite{Singh1,Singh2,SHChae16}, but with multiple contents ($M \gg 2$) analytic evaluation of \eqref{Rev:5_2} is hard due to the complicated form of $\bar{N}_i$.
To circumvent this difficulty, we again take a lower bound of \eqref{Rev:5_2} as
\begin{align}
\eqref{Rev:5_2}
\geq\sum_{i=1}^F f_i\cdot\mathbb{P}\left[\frac{1}{c}\log_2\left(1+\frac{P|h_{x_i}|^2|x_i|^{-\alpha}}{I_i}\right)\!\geq\!\rho_i\right],
\label{Rev:5_3}
\end{align}
where $c ~(\geq 1)$ is a constant independent of $i$ and
makes the inequality hold for all ranges of $\{p_i\}$, and $\tau_i=2^{c\rho_i}-1$.
Note that since \eqref{Rev:5_3} is a decreasing function with respect to $c$ and bounded below by
zero, there must exist a certain value of $c~(<\infty)$ which makes the inequality hold. The value of $c$ yielding a tight lower bound can be numerically determined; in general $c$ becomes larger as $\{\rho_i\}$ diminishes and $\gamma$ grows.
Fig. \ref{Fig:Approx_check} validates \eqref{Rev:5_3}, where
green and black dotted lines represent \eqref{Rev:5_2} and our lower bound in \eqref{Rev:5_3}, respectively.
It is verified that there exists a finite value of $c$ yielding a lower bound of \eqref{Rev:5_2}
regardless of $\{p_i\}$. In our setting, the value of $c$ for a tighter lower bound is $c\approx 40$.
Although there exists a gap between \eqref{Eqn:AVG_Ps_inter} and its lower bound \eqref{Rev:5_3},
the shape of those two functions looks quite similar and thus the caching probabilities obtained from \eqref{Rev:5_3}
are close to the optimal caching probabilities.

The equation \eqref{Rev:5_3} can be written by
\begin{align}
&\sum_{i=1}^F f_i\cdot\mathbb{P}\left[\frac{1}{c}\log_2\left(1+\frac{P|h_{x_i}|^2|x_i|^{-\alpha}}{I_i}\right)\geq\rho_i\right]\\
&=\sum_{i=1}^F f_i\cdot\mathbb{P}\left[
|h_{x_i}|^2\geq\frac{I_i\tau_i|x_i|^{\alpha}}{P}\right]\\
&\stackrel{(a)}{=}\sum_{i=1}^F f_i \int_0^{\infty}\!\mathbb{E}_{I_i}\!\left[\frac{\Gamma(m_D,m_DP^{-1}\tau_i r^{\alpha}I_i)}{\Gamma(m_D)}\right]\! f_{|x_i|}(r)dr,
\label{lower_AFTSP}
\end{align}
where $\tau_i=2^{c\rho_i}-1$, $\Gamma(s)$ is the Gamma function defined as $\Gamma(s)=\int_0^{\infty}t^{s-1}e^{-t}dt$, $\Gamma(s,x)$ is the upper incomplete Gamma function defined as $\Gamma(s,x)=\int_x^{\infty}t^{s-1}e^{-t}dt$,
$x_i$ is the location of the nearest caching helper storing content $i$,
$f_{|x_i|}(r) \left(=2\pi p_i\lambda r \exp\left(-\pi p_i\lambda r^2\right)\right)$ is the PDF of the distance to the nearest caching helper storing content $i$, and
\begin{align}
I_i
&= \sum_{y\in\Phi\setminus x_i} P|h_y|^2|y|^{-\alpha}\nonumber\\
&=\sum_{y\in\Phi_i^c} P|h_y|^2|y|^{-\alpha}+\sum_{y\in\Phi_i\setminus x_i} P|h_y|^2|y|^{-\alpha}.
\end{align}
The equality (a) is obtained from the Nakagami-$m_D$ fading channel power gain.

Since $\frac{\Gamma[m,my]}{\gamma(m)}=e^{-my}\sum_{k=0}^{m-1}\frac{m^k}{k!}y^k$, we have
\begin{align}
&\mathbb{E}_{I_i}\left[\frac{\Gamma(m_D,m_DP^{-1}\tau_i r^{\alpha}I_i)}{\Gamma(m_D)}\right]\\
&=\sum_{k=0}^{m_D-1}\frac{1}{k!}\left(m_D P^{-1}\tau_i r^{\alpha} \right)^k\mathbb{E}_{I_i}\left[I_i^ke^{-m_DP^{-1}\tau_i r^{\alpha}I_i}\right]\\
&\stackrel{(b)}{=}\!\sum_{k=0}^{m_D-1}\!\frac{1}{k!}\left(-m_D P^{-1}\tau_i r^{\alpha}\right)^k
\!\frac{d^k}{ds^k}\mathcal{L}_{I_i}(s)|_{s=\frac{m_D\tau_i r^{\alpha}}{P}}, \label{Inner}
\end{align}
where (b) is from $\mathcal{L}_{x^kf(x)}(s)=(-1)^k\frac{d^k\mathcal{L}_f(s)}{ds^k}$ and $\mathcal{L}_{I_i}(s)$ is the Laplace transform of $I_i$ given by
\begin{align}
&\mathcal{L}_{I_i}(s)
=\mathbb{E}\left[e^{-sI_i}\right]
= \mathbb{E}\left[e^{-s\sum_{y\in \Phi\setminus x_i}P|h_y|^2|y|^{-\alpha}}\right]\\
&\stackrel{(c)}{=}\mathbb{E}\left[\prod_{y\in \Phi\setminus x_i}
    \mathbb{E}_{|h_y|^2}\left[e^{-sP|h_y|^2|y|^{-\alpha}}\right]\right]\\
&\stackrel{(d)}{=} \exp\left(-2\pi p_i\lambda\int_{r}^{\infty}\left[1-\mathbb{E}_g\left[e^{-sPgv^{-\alpha}}\right]\right]vdv\right)\nonumber\\
&~~~\times\exp\!\left(\!-2\pi(1 \!-\! p_i)
\lambda\!\int_0^{\infty}\!\left[1\!-\!\mathbb{E}_g\!\left[e^{-sPgv^{-\alpha}}\right]\right]vdv\!\right)\\
&\stackrel{(e)}{=} \exp\left(-2\pi p_i\lambda\int_{r}^{\infty}\frac{(sPv^{-\alpha}+m_I)^{m_I}-m_I}{(sPv^{-\alpha}+m_I)^{m_I}}vdv\right)\nonumber\\
&~\times\exp\!\left(\!-2\pi(1\!-\!p_i)
\lambda\!\!\int_0^{\infty}\!\!\frac{(sPv^{-\alpha}\!+\! m_I)^{m_I}\!-\! m_I}{(sPv^{-\alpha}\!+\! m_I)^{m_I}}vdv\!\right)\\
&= \exp\left(-2\pi\lambda\int_0^{\infty}\frac{(sPv^{-\alpha}+m_I)^{m_I}-m_I}{(sPv^{-\alpha}+m_I)^{m_I}}vdv\right)\nonumber\\
&~~~\times\exp\left(2\pi p_i\lambda\int_0^r\frac{(sPv^{-\alpha}+m_I)^{m_I}-m_I}{(sPv^{-\alpha}+m_I)^{m_I}}vdv\right), \label{Laplace}
\end{align}
where (c) is due to independence of the channel;
(d) comes from the Probability Generating Functional (PGFL) of PPP; (e) is from the mogment generating function (MGF) of the Nakagami-$m_I$ distribution.

Substituting \eqref{Inner} into \eqref{lower_AFTSP},
we obtain
\begin{align}
P_s\geq \sum_{i=1}^F f_i \int_0^{\infty}
&\sum_{k=0}^{m_D-1}\frac{1}{k!}\left(-m_D P^{-1}\tau_i r^{\alpha}\right)^k\nonumber\\
&\times\frac{d^k}{ds^k}\mathcal{L}_{I_i}(s)|_{s=m_DP^{-1}\tau_i r^{\alpha}}f_{|x_i|}(r)dr,
\end{align}
where $m_D$ and $m_I$ are the Nakagami fading parameters of the desired and interfering links, respectively, and
\begin{align}
\mathcal{L}_{I_i}(s)
&=\exp\left(-2\pi\lambda\int_0^{\infty}\left[ 1 - \frac{m_I}{(sPv^{-\alpha}+ m_I)^{m_I}}\right]vdv\right.\nonumber\\
&\left.+2\pi p_i\lambda\!\int_0^r \left[ 1 - \frac{m_I}{(sPv^{-\alpha}+m_I)^{m_I}}\right]vdv\right),\\
f_{|x_i|}(r) &= 2\pi p_i\lambda r\exp\left(-\pi p_i\lambda r^2\right).
\end{align}


\end{document}